# SPATIAL-VARIANT CAUSAL BAYESIAN INFERENCE FOR RAPID SEISMIC GROUND FAILURES AND IMPACTS ESTIMATION

Xuechun Li[1], Susu Xu[2]

[1] Johns Hopkins University, Baltimore, USA, xli359@jhu.edu
[2] Johns Hopkins University, Baltimore, USA, sxu83@jhu.edu

**Abstract**: *Rapid and accurate estimation of post-earthquake ground failures and building damage is critical for effective post-disaster responses. Progression in remote sensing technologies has paved the way for the rapid acquisition of detailed, localized data, enabling swift hazard estimation through the analysis of correlation deviations between pre- and post-quake satellite imagery. However, discerning seismic hazards and their impacts is challenged by overlapping satellite signals from ground failures, building damage, and environmental noise. Previous advancements introduced a novel causal graph-based Bayesian network that continually refines seismic ground failure and building damage estimates derived from satellite imagery, accounting for the intricate interplay among geospatial elements, seismic activity, ground failures, building structures, damages, and satellite data. However, a notable shortcoming of this model is its neglect of spatial heterogeneity across different locations in a seismic region, which might curtail its precision in encapsulating the spatial diversity of seismic effects and repercussions.*

*In this study, we acknowledge the significance of spatial relationships in estimating post-earthquake scenarios and pioneer an approach that accounts for these spatial intricacies. Instead of relying solely on localized data for each hazard at a specific site, we introduce a spatial variable — influenced by the bilateral filter — to encapsulate the spatial relationships from surrounding hazards. The bilateral filter, an advanced computational tool, considers the spatial proximity of neighboring hazards, such as ground failures and building damage, and their ground shaking intensity values, ensuring refined modeling of spatial relationships. This integration captures a nuanced balance between individual site-specific hazard characteristics and overarching spatial tendencies, offering a comprehensive representation of the post-disaster landscape. Our model, tested across multiple earthquake events, underscores the significance of embracing spatial heterogeneity in seismic hazard and damage estimation. The results highlight a marked enhancement in post-earthquake large-scale multi-impact estimation accuracy and efficiency to effectively inform rapid disaster responses.*

## 1. Introduction

Due to their destructive nature, earthquakes not only cause direct building damage from ground shaking but also can trigger cascading ground failures like landslides and liquefaction. Along with ground shaking, these ground failures can have significant consequences, including building damage and loss of life. For example, the 2021 Haiti earthquake triggered 7,091 landslides over more than 80 km2 of land, which damaged or destroyed more than 130,000 buildings and led to 2,248 deaths with over 12,200 injured (Havenith et al., 2022; Web, 2021). Rapid assessment of the locations and severity of ground failures and building damage after an earthquake is crucial for the timely rescue of victims during the critical "Golden 72 Hour" period, as well as for subsequent disaster recovery strategies (Jang et al., 2009; Li et al., 2023).

Over the years, researchers have developed various approaches for estimating the location and intensity of earthquake-induced ground failures and building damage. Traditional methods encompass both physical models (Jibson, 2000; Marc et al., 2009; Newmark, 2023) and statistical counterparts (Jessee et al., 2016; Nowicki et al., 2014; Seed and Idriss, 1971; Zhu et al., 2015). Fundamental physical processes form the basis of physical models, exemplified by the Newmark displacement-centric landslide model (Jibson, 2000; Newmark, 2023) and the liquefaction potential index (Zhu et al., 2016). However, their application is limited when geotechnical data are absent. Despite their foundation in primary physical processes, these models can be prone to errors, often from oversimplified representations of intricate physical dynamics. Building fragility curve, a log-normal function that estimates the probability of different damage states from given seismic shaking and building types, is a traditional approach to estimate building damage. However, detailed building-type datasets are often unavailable in large-scale disaster zones, making fragility curves not scalable for rapid disaster impact estimation (FEMA, 2020; Li et al., 2023).

Alternatively, statistical models are calibrated by leveraging historical data from prior ground failures. They estimate potential failures using geospatial susceptibility indicators, such as slope and lithology, paired with approximated ground motion data that previously triggered similar failures (Jessee et al., 2016; Nowicki et al., 2014). The precision and granularity of these models, however, face restrictions due to constrained access to geospatial attributes and inherent modeling ambiguities. Furthermore, the intricate interplay between geospatial proxies, ground shaking, moisture conditions, and their combined influence on the likelihood of landslides and liquefaction is riddled with spatial relationships and significant uncertainties. An ongoing challenge lies in tailoring and extrapolating such models grounded in historical data to novel events. This is primarily because ground failures can be influenced by nuanced environmental elements that differ with each event.

Recent advances in remote sensing technologies also play a significant role in earthquake damage assessments. The NASA Advanced Rapid Imaging and Analysis (ARIA) team of researchers developed novel remote sensing methods that allow for rapid ground failure and damage estimates (Harp and Jibson, 2002; Lee, 2005; Zhao and Lu, 2018). Most notably, damage proxy maps (DPMs) harness multi-temporal variations between pre- and post-earthquake satellite imagery to pinpoint changes on the ground surface induced by earthquakes (Yun et al., 2015). DPMs are generated from synthetic aperture radar (SAR) data captured before and after radar reflection-changing events, such as earthquakes or landslides (Loos et al., 2020; Yun et al., 2015). Differences in satellite distance indicate variations in the ground surface, while shifts in spatial correlation suggest alterations in other surface elements, whether natural or anthropogenic (Burrows et al., 2021; Li et al., 2021; Zimmaro et al., 2020). However, imaging-based assessments face difficulties in distinguishing among specific types of changes, such as ground failures, building damages, and noise from vegetation growth and human-made alterations, especially when these alterations have spatial overlap (Kongar et al., 2017).

Recognizing the complexity of these challenges, researchers have turned to more advanced statistical methodologies, such as the causal graph-based Bayesian network (Xu et al., 2022; Xu et al., 2022; Li et al., 2023). This innovative approach employs a unique structure that perpetually refines the estimates of seismic ground failure and building damage using satellite images by considering the intertwined physical relationships between geospatial attributes, ground movements, structural damage, and satellite imagery. The limitation of this approach, however, is that it did not account for spatial heterogeneity across the different locations. Spatial heterogeneity refers to the variations in the patterns of earthquake-induced hazards or damage across different locations. For example, one area might experience a dense cluster of landslides due to its specific soil and geological attributes, while another region may remain largely unaffected because of its rocky terrain or because it is shielded by natural barriers such as a hill. The presence of such heterogeneity is also influenced by factors like local topography, land use, and subsurface conditions, which have been reported to play significant roles in determining ground response during earthquakes (Fan et al., 2019). The latest approach (Xu et al., 2022; Xu et al., 2022; Li et al., 2023) primarily focuses on direct causal relations and does not account for the complex nature of these spatial variables and their relationships. As a result, it tends to overlook the nuanced interplay of spatial relationships, leaving room for improvements in both accuracy and comprehensiveness. Recognizing these spatial relationships among neighboring areas is crucial for a more reliable estimation of post-earthquake damages.

To address the challenge of spatial heterogeneity in post-earthquake seismic hazard estimation, we have developed a new method to explore the spatial relationships among different locations. Instead of assuming each location to be independent, we studied how these locations may be related to their neighbors in terms of hazard probability and spatial distance. Specifically, we assumed that a location will have a higher chance of having similar hazard probabilities to its neighboring locations as opposed to locations that are far away. For example, if one location has sustained high landslide damage, the nearby areas will have higher probabilities of having experienced landslides compared to locations with greater distance. By integrating such spatial relationships into our model, we can provide a more comprehensive depiction of the seismic landscape, allowing our model to not only integrate data related to ground shaking, seismic ground failures, and impacts that are visible through satellite imagery but also factor in the complicated relationships of nearby locations in a causal graph.

To be able to integrate spatial relationships into our seismic hazard estimation model, we employ a method that allows for precise delineation of the boundaries of each location. Bilateral filter is an edge preserving tool used in image-processing whose purpose is to maintain sharp boundaries while reducing noise (Tomasi and Manduchi, 1998; Paris et al., 2009). The significance of the bilateral filter in seismic data analysis lies in its capacity to differentiate and preserve sharp intensity shifts, ensuring that the intricate patterns and nuances of seismic activity are accurately represented. By utilizing the strengths of bilateral filters, we enhance our model's ability to capture and represent spatial relationships more effectively.

In light of the evident spatial relationships among neighboring locations and their inherent dependencies, we have initiated research to harness the abilities of the bilateral filter in a novel approach. We aim to approximate seismic ground failures and impacts utilizing the Bayesian network, which has shown to be a powerful tool for deciphering complex casualties among multiple variables from a group of data (Jensen and Nielsen, 2007; Koller and Friedman, 2009; Wasserman, 2004), while paying special attention to spatial relationships in neighboring locations. Our new spatial-variant Bayesian network allows us to encode causal dependencies among ground shaking, seismic ground failures and impacts like building damage, ground surface changes captured by satellite images, as well as the spatial relationships between neighboring locations in a causal graph. Nodes in the Bayesian network symbolize the random variables, and their causal connections are depicted as directed edges. Through Bayesian updating, we can derive the posterior distributions of these random variables by understanding their conditional dependencies. Furthermore, we introduce a spatial variable for each hazard at each location to capture the spatial relationships among adjacent locations, allowing us to represent the patterns and extent of seismic ground failures and impacts more accurately.

This work has four main contributions: (1) Introducing a spatial variable for each hazard at every site, thereby capturing the spatial relationships of proximate hazards. (2) Implementing the bilateral filter to calculate a weighted average that is the inverse of the distance to adjacent hazards, thereby representing the more pronounced influence of hazards that are in closer proximity. (3) The holistic integration of individual hazard data with overarching spatial patterns, providing a balanced perspective that combines granular detail with macro trends. (4) A thorough assessment of our revamped methodology across various earthquake events to validate its adaptability and efficacy.

By infusing the attributes of the bilateral filter and the Bayesian network, this paper aspires to enhance the reliability and accuracy of earthquake-induced ground failures and building damage assessment on a large scale using InSAR imageries, presenting an avant-garde approach for optimizing post-disaster interventions and support. The structure of this paper unfolds as follows: Section 2 delves into the intricate details of our novel approach, elucidating the introduction of spatial variables, their integration within the causal graph-based Bayesian network, and the specifics of the stochastic variational inference, culminating in the refined algorithm of Bayesian updating with spatial considerations. In Section 3, we subject our methodology to rigorous evaluation through select case studies, dissecting its performance and potential implications.

## 2. Methodologies

In this section, we delve into our spatially variant causal Bayesian method specifically for joint seismic ground failures and impact estimation. Central to our approach is a causal Bayesian network that integrates ground failures, building damage, and remote sensing observations and places a pronounced emphasis on spatial interdependencies among neighboring locations. Following this, we outline a stochastic variational inference technique to derive the posteriors of ground failure and building damage, harnessing the strengths of remote sensing observations and pre-existing geospatial models. Eventually, we present the outline of our optimization framework designed to determine the optimal weights, which encapsulate the statistical relationships among various predictors, ground failure, building damage, and both environmental and anthropogenic noise, as well as remote sensing observations. In subsequent sections, the DPMs previously mentioned are used to represent our remote sensing observations.

**2.1. Spatial-Variant Causal Bayesian Network for Jointly Estimating Ground Failures and Impacts**

To jointly estimate ground failures and building damage, we first build a generalized causal graph in Figure 1 to represent the causal relationships among ground failures, building damage, sensing observations, and the spatial relations among the neighboring locations. In our work, we utilize DPMs as our sensing observations.

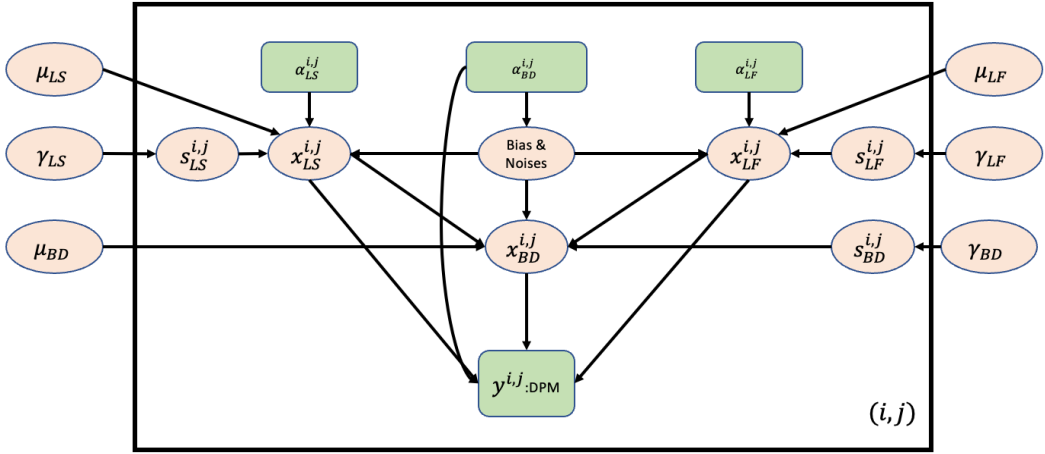

*Figure 1. Causal Spatial-Variant Bayesian Network for multi-hazards and impacts estimation in seismic events. $(i,j)$ in the feature refers to the location coordinate. Green rectangles refer to the known variables. Blue circles refer to unobserved nodes. $\gamma_h$ are the unknown causal parameters that quantify the causal relations among parent and child nodes. $\mu_h$ are the unknown parameters for the spatial variables.*

In a location with coordinate, $(i,j)$, we utilize use $y^{i,j}$, which is bounded by [0,1], to refer to our sensing observation. We denote $x_h^{i,j}$ to be the latent variable for ground failures and building damage, and $s_h^{i,j}$ to be the spatial variable, where we employ a bilateral filter to gauge the influence of neighboring hazards on a specific location, adjusting for both proximity and intensity to encapsulate spatial patterns in seismic data, where $h \in$ {Landslide ($LS$), Liquefaction ($LF$), Building Damage ($BD$)}. We assume the latent variable $x_h^{i,j}$ is a binary variable with a Bernoulli distribution. We also define $\epsilon_h^{i,j}$ to represent the term of environmental noises for node $x_h^{i,j}$, and $\epsilon_y^{i,j}$ the noise for node $y_h^{i,j}$.

Given sensing observations, $y^{i,j}$, we utilize $\mathcal{P}(y)$ to define the parents of $y$. We denote and $w_{k_i}$ to quantify the causal effects from a parent node $k$ to a child node $i$, where $k$ is any parent node of $i$. All weight nodes $w_{k_i}$ are defined as deterministic variables. We assume transformations from the parent nodes of sensing observations to $y$ are often modeled as a log-normal (LN) distribution (Xu et al., 2022; Li et al., 2023):

$$y^{i,j}|\mathcal{P}(y^{i,j}), \epsilon_{y^{i,j}} \sim LN(\sum_{k \in \mathcal{P}(y^{i,j})} w_{ky} x_k^{i,j} + w_{\epsilon_y} \epsilon_y^{i,j} + w_{0y}, w_{\epsilon_y}^2) \qquad (1)$$

For ground failures and building damage, we assume the latent variables, $x_h^{i,j}$, have values {0,1}. When LS, LF, or BD happens in location $(i,j)$, $x_h^{i,j} = 1$, otherwise, $x_h^{i,j} = 0$. We define a leak node, with index 0, that is always active. It allows its child nodes to be active even when other parent nodes are inactive. We use $\mathcal{P}(h^{i,j})$ to represent the parents of node $x_h^{i,j}$. The node activation probabilities are defined as follows:

$$p(x_h^{i,j}|\mathcal{P}(h^{i,j}),\epsilon_h^{i,j}) = \left[p(x_h^{i,j}=1|\mathcal{P}(h^{i,j}),\epsilon_h^{i,j})\right]^{x_h^{i,j}} \left[1 - p(x_h^{i,j}=1|\mathcal{P}(h^{i,j}),\epsilon_h^{i,j})\right]^{1-x_h^{i,j}}$$

$$p(x_h^{i,j}=1|\mathcal{P}(h^{i,j}),\epsilon_h^{i,j}) = \frac{1}{1+exp(-\sum_{k\in\mathcal{P}(h^{i,j})} w_{kh} x_k^{i,j} - w_{\epsilon_h}\epsilon_h^{i,j} - w_{0h} - w_{s_h} s_h^{i,j})} \quad (2)$$

For each hazard $h$ at each location $(i,j)$, we introduce a spatial variable $s_h^{i,j}$. This variable will represent the spatial effect of hazard $h$ at location $(i,j)$. We define $s_h^{i,j}$ to be influenced by the hazard $h$ at neighboring locations with coordinates $\{(i-1,j-1), (i,j-1), (i+1,j-1), (i-1,j), (i+1,j), (i-1,j+1), (i,j+1), (i+1,j+1)\}$. To capture the spatial relations among neighboring locations, we employ a bilateral filter to measure the impact of nearby hazards on a given location, considering both distance and intensity to capture the spatial intricacies within seismic data. The spatial variable is defined to be:

$$s_h^{i,j} = \frac{1}{M_h^{i,j}} \sum_{(i',j')\in\Omega_{(i,j)}} x_h^{i,j} \times f_r(\Delta x_{i',j',i,j}) \times g_s(||(i',j') - (i,j)||) \quad (3)$$

where $x_h^{i',j'}$ denotes the neighboring locations in the local neighborhood $\Omega_{i,j}$ around $x_h^{i,j}$. For each $x_h^{i',j'}$ in $\Omega_{i,j}$, we consider its intensity value $I(x_h^{i',j'})$, its spatial distance from $(i,j)$, and the intensity difference between $I(x_h^{i',j'})$ and $I(x_h^{i,j})$ to compute the filter response. $f_r(\Delta x_{i',j',i,j})$ is the range kernel, which measures the similarity in intensities between the central location $I(x_h^{i,j})$ and the neighboring location $I(x_h^{i',j'})$. It is modeled as a Gaussian function, $f_r(\Delta x_{i',j',i,j}) = e^{-\frac{(\Delta x_{i',j',i,j})^2}{2\sigma_r^2}}$, where large intensity differences result in small weights. Similarly, $g_s(||(i',j') - (i,j)||) = e^{-\frac{(||(i',j')-(i,j)||)^2}{2\sigma_s^2}}$, is the spatial kernel, which measures the spatial closeness between the central pixel location $(i,j)$ and the neighboring location $(i',j')$. It is also modeled as a Gaussian function. $M_h^{i,j}$ is a normalization term to ensure the sum of the weights to 1. It is computed as the sum of all the weights given by the product of the range and spatial kernels for all $x_h^{i',j'}$ in $\Omega_{(i,j)}$.

Utilizing the above distributions and conditional distribution assumptions, we develop a Bayesian network rooted in the causal graph, adeptly representing the interdependencies among ground failures, building damage, remote sensing observations, and spatial relations. However, the intricate statistical dependencies make the posterior of unobserved variables for ground failures and building damage intractable. To address this, we introduce a stochastic variational inference framework designed to approximate the otherwise elusive posterior concerning unobserved ground failure and building damage.

## 2.2. Variational Inference for Posteriors Approximation

With the causal spatial-variant Bayesian network constructed in Section 2.1, we further develop a variational inference algorithm to factorize the Bayesian network and jointly estimate the posterior distributions of latent variables and causal dependencies. We aim to jointly infer the posteriors of the latent variables in the Bayesian network, which represents the target ground failures, building damage, and spatial variables, considering the spatial relations among neighboring locations with unknown parameters of causal dependencies. For each location with coordinate $(i,j)$, we define $q_h^{i,j}$ to be the approximate posterior probability that hazard $h$ at location $(i,j)$ is active. Then, the variational distribution $q(X^{i,j})$ that factorizes over hidden nodes is defined to be:

$$q(X^{i,j}) = \prod_h (q_h^{i,j})^{x_h^{i,j}} (1 - q_h^{i,j})^{1-x_h^{i,j}} \quad (4)$$

We can then derive a tight lower bound on the marginal likelihood of the observed variable $y$ as (Jordan et al., 1999):

$$log\, p(Y) = \sum_{(i,j)} log \int p(y^{i,j}, S^{i,j}, X^{i,j}, \epsilon^{i,j})\, d(X^{i,j}, \epsilon^{i,j}) \qquad (5)$$

$$\geq \mathbb{E}_{q(X^{i,j},\epsilon^{i,j})}[log\, p(y^{i,j}, S^{i,j}, X^{i,j}, \epsilon^{i,j})] - \mathbb{E}_{q(X^{i,j},\epsilon^{i,j})}[log\, q(X^{i,j}, \epsilon^{i,j})]$$

To obtain the final lower bound, we can further derive the first term, $\mathbb{E}_{q(X^{i,j},\epsilon^{i,j})}[log\, p(y^{i,j}, S^{i,j}, X^{i,j}, \epsilon^{i,j})]$, in Equation 5 as:

$$\mathbb{E}_{q(X^{i,j},\epsilon^{i,j})}[log\, p(y^{i,j}, S^{i,j}, X^{i,j}, \epsilon^{i,j})]$$

$$= \mathbb{E}[log\, p(y^{i,j}|\mathcal{P}(y^{i,j}), \epsilon_y^{i,j})] + \sum_h \mathbb{E}(log\, p(x_h^{i,j}|\mathcal{P}(x_h^{i,j}), \epsilon_h^{i,j})) + \sum_h \mathbb{E}(log\, p(s_h^{i,j}))$$

$$+ \sum_h \mathbb{E}(log\, p(\epsilon_h^{i,j})) + \mathbb{E}(log\, p(\epsilon_y^{i,j})) \qquad (6)$$

where:

$$\mathbb{E}[p(y^{i,j}|\mathcal{P}(y^{i,j}), \epsilon_y^{i,j})]$$

$$= -log\, y^{i,j} - log\, |w_{\epsilon_y}| - \frac{\sum_{m \neq n} w_{my} w_{ny} q_m^{i,j} q_n^{i,j}}{w_{\epsilon_y}^2} - \frac{(log\, y)^2 + w_{0y}^2 + \sum_{k \in \mathcal{P}(y^{i,j})} w_{ky}^2 q_k^{i,j}}{2 w_{\epsilon_y}^2}$$

$$- \frac{(w_{0y} - log\, y^{i,j}) \sum_{k \in \mathcal{P}(y^{i,j})} w_{ky} q_k^{i,j} - w_{0y} log\, y^{i,j}}{w_{\epsilon_y}^2} \qquad (7)$$

According to Equation 2, for each hidden node $x_h^{i,j}$, where $h \in \{LS, LF, BD\}$, we can derive the expectation of the log-likelihood, $\mathbb{E}(log\, p(x_h^{i,j}|\mathcal{P}(x_h^{i,j}), \epsilon_h^{i,j}))$, as:

$$\mathbb{E}(log\, p(x_h^{i,j}|\mathcal{P}(x_h^{i,j}), \epsilon_h^{i,j})) = q_h^{i,j} \mathbb{E}\left(-log\left(1 + exp\left(-\sum_{k \in \mathcal{P}(h^{i,j})} w_{kh} x_k^{i,j} - w_{\epsilon_h} \epsilon_h^{i,j} - w_{0h} - w_{s_h} s_h^{i,j}\right)\right)\right)$$

$$+ (1 - q_h^{i,j}) \mathbb{E}\left(-log(1 + exp\left(\sum_{k \in \mathcal{P}(h^{i,j})} w_{kh} x_k^{i,j} + w_{\epsilon_h} \epsilon_h^{i,j} + w_{0h} + w_{s_h} s_h^{i,j}\right))\right) \qquad (8)$$

where the distribution of log-sum-exp term in Equation 8 is intractable. Therefore, we need to get a tight lower bound of the expectation. By Jensen's inequality and Taylor's theorem, we obtain:

$$\mathbb{E}(-log(1 + exp\, t)) \geq -log(1 + \mathbb{E}(exp\, t)) \qquad (9)$$

The lower bound of $\mathbb{E}(log\, p(x_h^{i,j}|\mathcal{P}(x_h^{i,j}), \epsilon_h^{i,j}))$ in Equation 6 is derived as:

$$\mathbb{E}(log\, p(x_h^{i,j}|\mathcal{P}(x_h^{i,j}), \epsilon_h^{i,j}))$$

$$\geq -q_h^{i,j} log\left(1 + exp\left(-w_{0h} + \frac{w_{\epsilon_h}^2}{2} - w_{s_h} s_h^{i,j}\right)\right) \left(exp\left(-\sum_{k \in \mathcal{P}(h^{i,j})} w_{kh}\right) q_k^{i,j} + (1 - q_k^{i,j})\right)$$

$$-(1 - q_h^{i,j}) log\left(1 + exp\left(w_{0h} + \frac{w_{\epsilon_h}^2}{2} + w_{s_h} s_h^{i,j}\right)\right) \left(exp\left(\sum_{k \in \mathcal{P}(h^{i,j})} w_{kh}\right) q_k^{i,j} + (1 - q_k^{i,j})\right) \qquad (10)$$

We employ Monte Carlo estimates to evaluate $\mathbb{E}(log\, p(s_h^{i,j}))$, where $s_h^{i,j}$ derived from neighboring values of $x_h^{i',j'}$, a characteristic of the bilateral filter. Given the dependency of $s_h^{i,j}$ on its neighbors, our first step is to sample from $p(x_h)$. To do this, we draw $N$ samples, denoted as $x_h^n$, from $p(x_h)$, where $n$ represents the sample number, and $x_h^n$ signifies the complete spatial field for the $n^{th}$ sample. For each spatial sample $x_h^n$, we calculate the value of $s_{h,n}^{i,j}$ using the bilateral, factoring in the defined neighboring values at each $(i,j)$ location. Subsequently, we derive $log\, p(s_{h,n}^{i,j})$ from the empirical distribution. As $x$ is a binary discrete variable, we tally the occurrence frequency of each distinct value of $s_{h,n}^{i,j}$ within the dataset. The probability $p(s_{h,n}^{i,j})$ is then the frequency of a specific value divided by the total number of data points $N$. The Monte Carlo estimate is then computed as the average of the log probabilities, that is, $\mathbb{E}(log\, p(s_h^{i,j})) = \frac{1}{N}\sum_n p(s_{h,n}^{i,j})$.

Given a map containing a set of locations with coordinates $(i,j)$, we can therefore derive a tight variational lower bound as follows:

$$\mathcal{L}(q,w) = \sum_{(i,j)} \{-log\, y^{i,j} - log\, |w_{\epsilon_y}| - \frac{\sum_{m\neq n} w_{my} w_{ny} q_m^{i,j} q_n^{i,j}}{w_{\epsilon_y}^2} - \frac{(log\, y)^2 + w_{0y}^2 + \sum_{k\in\mathcal{P}(y^{i,j})} w_{ky}^2 q_k^{i,j}}{2w_{\epsilon_y}^2}$$

$$- \frac{(w_{0y} - log\, y^{i,j})\sum_{k\in\mathcal{P}(y^{i,j})} w_{ky} q_k^{i,j} - w_{0y}\, log\, y^{i,j}}{w_{\epsilon_y}^2}$$

$$+ \sum_{h\in\{LS,LF,BD\}} \left\{ -q_h^{i,j}\, log\left(1 + exp\left(-w_{0h} + \frac{w_{\epsilon_h}^2}{2} - w_{s_h} s_h^{i,j}\right)\right)\left(exp\left(-\sum_{k\in\mathcal{P}(h^{i,j})} w_{kh}\right) q_k^{i,j} + (1 - q_k^{i,j})\right) - (1$$

$$- q_h^{i,j})\, log\left(1 + exp\left(w_{0h} + \frac{w_{\epsilon_h}^2}{2} + w_{s_h} s_h^{i,j}\right)\right)\left(exp\left(\sum_{k\in\mathcal{P}(h^{i,j})} w_{kh}\right) q_k^{i,j} + (1 - q_k^{i,j})\right) \right\}$$

$$+ \frac{1}{N}\sum_n \sum_h p(s_{h,n}^{i,j}) - \sum_{h\in\{LS,LF,BD\}} [q_h^{i,j} log\, q_h^{i,j} + (1 - q_h^{i,j}) log(1 - q_h^{i,j})]\}$$

(11)

### 2.2. Stochastic Optimization

We aim to minimize the loss function, targeting the optimal melding of posteriors and the parameters governing causal dependencies. To achieve this, we employ an expectation-maximization (E-M) algorithm designed to iteratively optimize both the posteriors and causal parameters. Each iteration begins with the selection of a mini-batch of locations, after which the expectation and maximization steps are alternated. The expectation and maximization steps are executed to refine the posterior estimates and adjust the global weight parameters. Upon convergence of the model, we derive the optimal posteriors for landslide, liquefaction, and building damage at every specified location. As for the maximization step, we use stochastic gradient updates on the weights, leveraging the data from the sampled mini-batch. The weight updates for iteration t+1 are computed as follows:

$$w^{t+1} = w^t + \rho A \nabla \mathcal{L}(w) \quad (12)$$

where we utilize a positive definite preconditioner, denoted as $A$, along with a specific learning rate $\rho$, we update the model based on the gradient scheme. The gradients $\nabla\mathcal{L}(w)$ correspond to the loss function with respect to the weights. Under certain conditions detailed in (Robbins and Monro, 1951), this gradient update approach ensures convergence to a local maximum of $\mathcal{L}$. Upon model convergence, we secure the optimal posteriors for LS, LF, and BD at every location.

## 3. Evaluations

In this section, we evaluate our spatial-variant causal Bayesian inference method for rapid seismic ground failures, which are landslides and liquefaction in our case, and building damage estimation using two events: the 2020 Puerto Rico and the 2021 Haiti earthquakes. We utilize the current USGS ground failure products for landslides and liquefaction, providing individual seismic hazard estimations as our prior ground failure models (Allstadt et al., 2017; Nowicki et al., 2018; Zhu et al., 2015). We also use building fragility curves, which are log-normal functions that estimate the probability of different damage states, given seismic shaking and building types, as the prior model for building damage (FEMA, 2020). We benchmark against the prior models and, as well as the VCBI model detailed in (Xu et al., 2022).

We utilize the receiver operating characteristics (ROC) curve and the area under the ROC curve (AUC) as the evaluation metrics. The ROC curve is a graphical representation that illustrates the diagnostic ability of a binary classifier system as its discrimination threshold varies. It plots the True Positive Rate (sensitivity) against the False Positive Rate (1-specificity) for different threshold values. The Area Under the Curve (AUC) represents the degree or measure of separability, indicating how well the model distinguishes between positive and negative classes. A higher AUC value suggests better classifier performance, with a value of 1 indicating perfect classification and a value of 0.5 denoting no discrimination capability.

**3.1. The evaluation results of the 2020 Puerto Rico earthquake**

On January 7, 2020, the southwest region of Puerto Rico experienced an earthquake with a magnitude of Mw 6.4. Following this seismic event, reports indicated that the epicenter's vicinity triggered at least 300 landslides (Allstadt et al., 2022). In response to the earthquake, the ARIA team generated DPMs utilizing SAR images from the Sentinel-1 satellite (ARIA, 2020). Subsequently, a comprehensive field reconnaissance was carried out by specialists from USGS, University of Puerto Rico Mayagüez, the GEER team, and the StEER team to gather ground truth observations (Günay et al., 2020; Miranda et al., 2020).

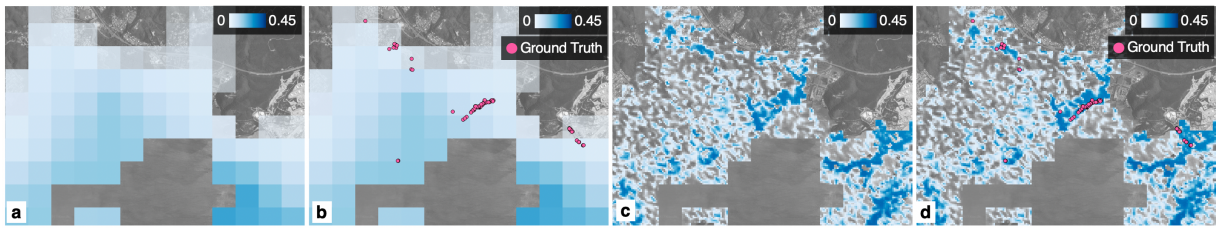

*Figure 2. Geospatial prior and posterior liquefaction estimation models of the 2020 Puerto Rico earthquake. (a) Prior liquefaction model. (b) Prior liquefaction model with ground truth observations. (c) Posterior liquefaction model. (d) Posterior liquefaction model with ground truth observations.*

Evaluation findings indicate that the posterior liquefaction model, depicted in Figure 2c, aligns more closely with the spatial distribution of ground truth observations shown in Figure 2d compared to the prior models illustrated in Figure 2a. By incorporating the DPM, prior models, and the spatial interconnections of neighboring locations, our model successfully detects a more significant number of liquefaction occurrences that align with ground truth observations, outperforming the existing USGS models. The ROC curves of our model and the baseline models are shown in Figure 4a. The AUC value of our posterior liquefaction model, which is 0.9451, demonstrates that our model outperforms the baseline USGS liquefaction model (AUC: 0.8662) and the VCBI posterior liquefaction model (AUC: 0.9121).

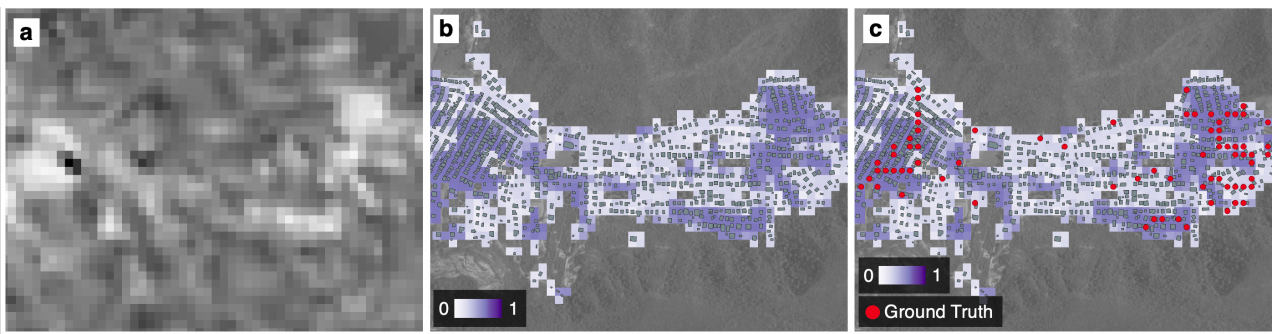

*Figure 3. Geospatial prior and posterior building damage estimation models of the 2020 Puerto Rico earthquake. (a) DPM with significant signal changes. (b) posterior building damage model with building footprints. (c) posterior building damage model with building footprints and ground truth observations.*

Figure 3 illustrates the performance of our model in estimating building damage in the quake-affected region, juxtaposed with building footprints and actual ground truth observations. The regions with confirmed building damage as per ground truth observations correspond closely with areas where our model predicts a high likelihood of damage. The AUC metrics highlight the superior performance of our posterior building damage models compared the baseline prior building damage model and the VCBI model. We present the ROC curves for the two baseline models and our model in Figure 4b. With an AUC value of 0.9720, our model exhibits a 13.7% enhancement over the baseline prior building damage model, which stands at 0.8550 and outperforms the baseline VCBI model with an AUC of 0.9309.

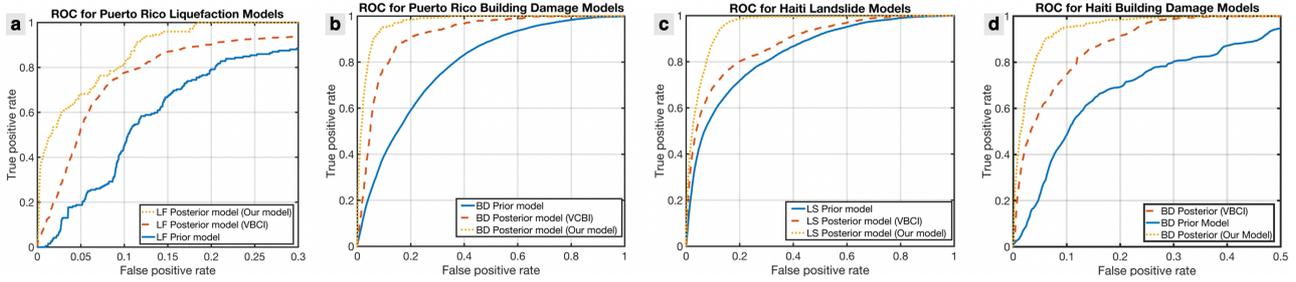

*Figure 4. ROC curve of baseline models and our posterior model for (a) The 2020 Puerto Rico liquefaction estimation. (b) The 2020 Puerto Rico building damage estimation. (c) The 2021 Haiti landslide estimation. (d) The 2021 Haiti building damage estimation.*

**3.2. The evaluation results of the 2021 Haiti earthquake.**

On August 14, 2021, a significant earthquake measuring Magnitude 7.2 struck the southern peninsula of Haiti. Subsequent post-disaster evaluations revealed that this seismic event destroyed 53,815 buildings and damaged an additional 83,770 in Grand Anse, Nippes, and Sud regions. The tragedy also resulted in at least 2,248 fatalities (USGS, 2021; Web, 2021). Both the StEER and GEER teams undertook dedicated efforts to gather ground truth data on landslides (LS) and building damage (BD) (GEER, 2021; Kijewski-Correa et al., 2021; Zhao et al, 2021).

Our evaluation results highlight the enhanced accuracy of our posterior model, as visualized in Figure 5c, especially when juxtaposed against the ground truth observations in Figure 5d. Compared to the prior models in Figure 5a, our higher-resolution landslide model exhibits a more faithful representation of observed landslide occurrences. This precision is underscored by its performance against existing USGS models. Furthermore, the ROC curves for the three models are shown in Figure 4c. When gauged via the AUC metric, our posterior landslide model boasts a value of 0.9373, which is significantly superior to both the baseline USGS landslide model (with an AUC of 0.8951) and the VCBI posterior landslide model (AUC at 0.9032).

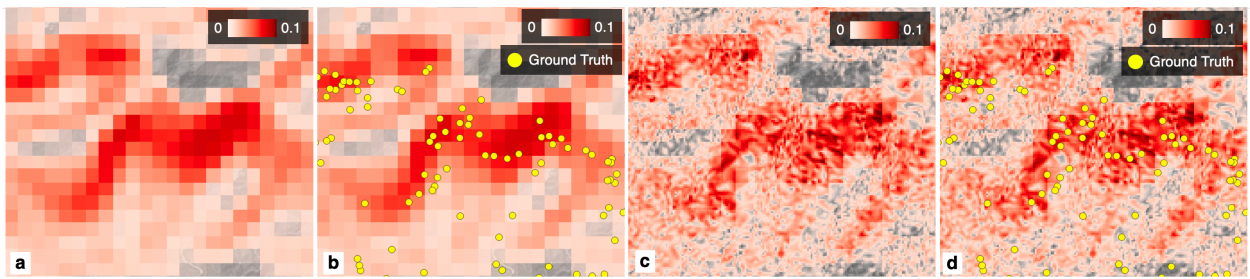

*Figure 5. Geospatial prior and posterior landslide estimation models of the 2021 Haiti earthquake. (a) Prior landslide model. (b) Prior landslide model with ground truth observations. (c) Posterior landslide model. (d) Posterior landslide model with ground truth observations.*

In the context of the 2021 Haiti earthquake, Figure 6 compares our model's proficiency in estimating building damage against DPMs, building footprints, and actual ground truth observations. A deeper dive into the visualization reveals our posterior model's tendency to assign a heightened probability to locations registering high values on the DPM. As per ground truth data, a congruence emerges between regions confirmed to have sustained building damage and those our model earmarks with a high probability of damage. The robustness of our posterior building damage models is further validated by ROC curves, which are presented in Figure 4d, and AUC metrics, distinctly outpacing the baseline VCBI model. Clocking an AUC of 0.9425, our model surpasses the VCBI baseline, pegged at 0.9021. Our model also makes an improvement margin of 14.65% over the baseline prior building damage model, which stands at 0.8220.

## 4. Conclusion

This study emphasized the fundamental importance of spatial relationships in post-earthquake hazard and damage estimation. Incorporating the bilateral filter into our model represents a significant leap forward in understanding and representing seismic aftermath. This tool, adept at deciphering localized nuances and overarching spatial trends, ensures a holistic representation of the affected landscape. By considering the spatial proximity and intensity of neighboring hazards, we achieve a granular and comprehensive understanding that goes beyond individual hazard characteristics. The effectiveness of our approach has been substantiated through its application to multiple earthquake events, each underscoring the robustness and adaptability of the model. With the ability to discern the intricacies of overlapping signals in satellite imagery—from ground failures, building damages, or environmental interferences—our model stands out as a beacon of precision and clarity. As we move forward, the findings from this research not only promise enhanced accuracy in post-earthquake analysis but also set the stage for shaping more efficient, strategic, and informed disaster recovery and emergency response protocols.

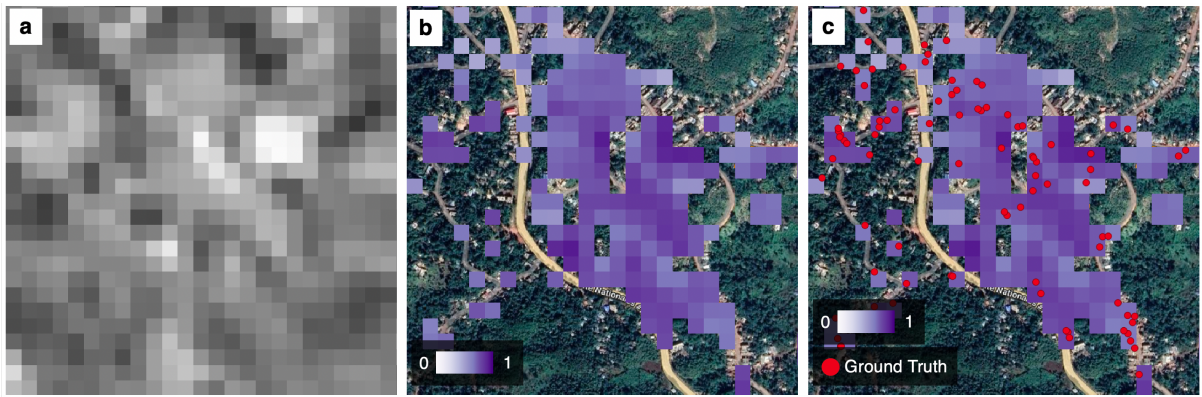

*Figure 6. Geospatial prior and posterior building damage estimation models of the 2021 Haiti earthquake. (a) DPM with significant signal changes. (b) posterior building damage model with building footprints. (c) posterior building damage model with building footprints and ground truth observations.*

## 5. Acknowledges

X. L. and S. X. are supported by U.S. Geological Survey Grant G22AP00032. This research is also supported by Stony Brook University and Stanford University through the resources provided. We would like to thank the support of Dr. Sang-ho Yun and Dr. Eric Fielding from NASA ARIA team for providing high-quality updated DPM products, and Dr. Kate Allstadt and Dr. Eric Thompson of the USGS National Earthquake Information Center and Jesse Rozelle of the Federal Emergency Management Agency (FEMA) contributed geospatial modeling and field observations. Any use of trade, firm, or product names is for descriptive purposes only and does not imply endorsement by the U.S. Government.